\documentclass[a4paper,USenglish,cleveref,autoref,thm-restate]{lipics-v2021}

\pdfoutput=1
\hideLIPIcs

\title{Higher-Order LCTRSs and Their Termination\footnote{This paper was presented at HOR 2023 on July 4 in Rome, Italy.}}

\author{Liye Guo}{Radboud University, Netherlands \and \url{https://www.cs.ru.nl/~liyeguo}}{l.guo@cs.ru.nl}{}{}
\author{Cynthia Kop}{Radboud University, Netherlands \and \url{https://www.cs.ru.nl/~cynthiakop/}}{c.kop@cs.ru.nl}{}{}
\authorrunning{L. Guo and C. Kop}
\Copyright{Liye Guo and Cynthia Kop}

\ccsdesc{Theory of computation~Equational logic and rewriting}
\keywords{Higher-order term rewriting, termination, recursive path ordering}

\funding{The authors are supported by the NWO VIDI project ``CHORPE'', NWO VI.Vidi.193.075.}

\nolinenumbers

\EventEditors{John Q. Open and Joan R. Access}
\EventNoEds{2}
\EventLongTitle{42nd Conference on Very Important Topics (CVIT 2016)}
\EventShortTitle{CVIT 2016}
\EventAcronym{CVIT}
\EventYear{2016}
\EventDate{December 24--27, 2016}
\EventLocation{Little Whinging, United Kingdom}
\EventLogo{}
\SeriesVolume{42}
\ArticleNo{23}

\usepackage{mathtools,amssymb}

\newcommand{\setBra}[1]{\left\{\,#1\,\right\}}
\newcommand{\setThe}[1]{{#1}_\vartheta}
\newcommand{\setSrt}{\mathcal{S}}
\newcommand{\setTyp}{\mathcal{T}}
\newcommand{\setFun}{\mathcal{F}}
\newcommand{\setVar}{\mathcal{V}}
\newcommand{\setPtm}{\mathbb{T}}
\newcommand{\setFva}[1]{\mathrm{Var}({#1})}
\newcommand{\setAlg}{\mathfrak{X}}
\newcommand{\setRul}{\mathcal{R}}

\newcommand{\arrTyp}{\to}
\newcommand{\arrRul}{\to}
\newcommand{\arrRwt}[1]{\arrRul_{#1}}
\newcommand{\arrJoi}[1]{\downarrow_{#1}}

\newcommand{\mapInt}[1]{[\![#1]\!]}
\newcommand{\mapSta}{\mathsf{stat}}

\newcommand{\relCom}[2]{{#1}\mathrel{;}{#2}}
\newcommand{\relCwf}{\succ}
\newcommand{\relQua}{\succsim}
\newcommand{\relTgt}{\sqsupset}
\newcommand{\relTge}{\sqsupseteq}
\newcommand{\relPre}{\blacktriangleright}
\newcommand{\relRpo}[4]{{#2}#1_{#4}{#3}}
\newcommand{\relUnt}{\triangleright}

\newcommand{\lblLex}{\mathfrak{l}}
\newcommand{\lblMul}{\mathfrak{m}}

\newcommand{\bSrt}{\mathbb{B}}
\newcommand{\bFal}{\bot}
\newcommand{\bTru}{\top}
\newcommand{\bInF}{\mathfrak{0}}
\newcommand{\bInT}{\mathfrak{1}}

\newcommand{\appl}[2]{{#1}\ {#2}}
\newcommand{\type}[2]{{#1}:{#2}}
\newcommand{\rwtr}[3]{{#1}\arrRul{#2}\ [{#3}]}

\newcommand{\trmA}{t}
\newcommand{\trmB}{s}
\newcommand{\trmC}{r}
\newcommand{\typA}{A}
\newcommand{\typB}{B}
\newcommand{\funA}{f}
\newcommand{\funB}{g}
\newcommand{\varA}{x}
\newcommand{\subA}{\sigma}
\newcommand{\rulL}{\ell}
\newcommand{\rulR}{r}
\newcommand{\rulC}{\varphi}
\newcommand{\valA}{v}

\newcommand{\init}{\mathsf{init}}
\newcommand{\exit}{\mathsf{exit}}
\newcommand{\comp}{\mathsf{comp}}
\newcommand{\fact}{\mathsf{fact}}

\begin{document}

\maketitle

\begin{abstract}
  Logically constrained term rewriting systems (LCTRSs) are a program analyzing formalism with native support for data types which are not (co)inductively defined.
  As a first-order formalism, LCTRSs have accommodated only analysis of imperative programs so far.
  In this paper, we present a higher-order variant of the LCTRS formalism, which can be used to analyze functional programs.
  Then we study the termination problem and define a higher-order recursive path ordering (HORPO) for this new formalism.
\end{abstract}

\section{Introduction}
\emph{Logically constrained term rewriting systems} (LCTRSs) \cite{kop:nis:13,fuh:kop:nis:17}
are a formalism for analyzing programs.
In real-world programming, data types such as
integers, as opposed to natural numbers, and arrays are prevalent.
Any practical technique for program analysis should be prepared to handle these.
One of the defining features of the LCTRS formalism
is its native support for such data types,
which are not (co)inductively defined and
need to be encoded if handled by more traditional TRSs.
Another benefit of the formalism is its separation between
logical constraints modeling the control flow and
other terms representing the program states.
So far, program analysis on the basis of LCTRSs has concerned imperative programs
since LCTRSs were introduced as a first-order formalism.
We are naturally curious to see
if functional programs can also be analyzed by constrained rewriting.
What we present here is our ongoing exploration in this direction:
First, we define a higher-order variant of the LCTRS formalism,
which, despite the absence of lambda abstractions,
is capable of representing some real-world functional programs straightforwardly.
Then we approach the termination problem for this new formalism
by defining a higher-order recursive path ordering (HORPO).
%

\section{LCSTRS}
We start defining
\emph{logically constrained simply-typed term rewriting systems} (LCSTRSs)
with types and terms.
We postulate a set \( \setSrt \),
whose elements we call \emph{sorts},
and a subset \( \setThe{\setSrt} \) of \( \setSrt \),
whose elements we call \emph{theory sorts}.
The set \( \setTyp \) of \emph{types} and
its subset \( \setThe{\setTyp} \),
called the set of \emph{theory types},
are generated as follows:
\( \setTyp          \Coloneqq \setSrt          \mid (\setTyp          \arrTyp \setTyp) \) and
\( \setThe{\setTyp} \Coloneqq \setThe{\setSrt} \mid (\setThe{\setSrt} \arrTyp \setThe{\setTyp}) \).
Right-associativity is assigned to \( \arrTyp \)
so we can omit some parentheses in types.
We assume given disjoint sets \( \setFun \) and \( \setVar \),
whose elements we call \emph{function symbols} and \emph{variables}, respectively.
The grammar
\( \setPtm \Coloneqq \setFun \mid \setVar \mid (\appl{\setPtm}{\setPtm}) \)
generates the set \( \setPtm \) of \emph{pre-terms}.
Left-associativity is assigned to the juxtaposition operation in the above grammar
so \( \appl{\appl{\trmA_0}{\trmA_1}}{\trmA_2} \) stands for
\( (\appl{(\appl{\trmA_0}{\trmA_1})}{\trmA_2}) \), for example.
We assume that every function symbol and variable is assigned a unique type.
Typing works as expected:\ %
if pre-terms \( \trmA_0 \) and \( \trmA_1 \)
have types \( \typA \arrTyp \typB \) and \( \typA \), respectively,
\( \appl{\trmA_0}{\trmA_1} \) has type \( \typB \).
Pre-terms having a type are called \emph{terms}.
We write \( \type{\trmA}{\typA} \) if a term \( \trmA \) has type \( \typA \).
We postulate a subset \( \setThe{\setFun} \) of \( \setFun \),
whose elements we call \emph{theory (function) symbols},
and assume that theory symbols have theory types.
Terms constructed with only theory symbols and variables
are called \emph{theory terms}.
The set of variables in a term \( \trmA \),
denoted by \( \setFva{\trmA} \), is defined as follows:
\( \setFva{\funA} = \emptyset \),
\( \setFva{\varA} = \setBra{\varA} \) and
\( \setFva{\appl{\trmA_0}{\trmA_1}} = \setFva{\trmA_0} \cup \setFva{\trmA_1} \).
A term \( \trmA \) is called a \emph{ground term} if \( \setFva{\trmA} = \emptyset \).
Note that ground theory terms always have theory types.
Theory terms are distinguished
because they will be treated specially when we define the rewrite relation.
First, let us define the interpretation of ground theory terms.
We postulate an \( \setThe{\setSrt} \)-indexed family of sets
\( (\setAlg_\typA)_{\typA \in \setThe{\setSrt}} \),
and extend it to a \( \setThe{\setTyp} \)-indexed family of sets
by letting \( \setAlg_{\typA \arrTyp \typB} \) be the set of maps from \( \setAlg_\typA \) to \( \setAlg_\typB \).
Now we assume given a \( \setThe{\setTyp} \)-indexed family of maps \( (\mapInt{\cdot}_\typA)_{\typA \in \setThe{\setTyp}} \)
where \( \mapInt{\cdot}_\typA \)
assigns to each theory symbol whose type is \( \typA \) an element of \( \setAlg_\typA \) and
is bijective if \( \typA \in \setThe{\setSrt} \).
Theory symbols whose type is a theory sort are called \emph{values}.
We extend each indexed map \( \mapInt{\cdot}_\typB \) to a map
that assigns to each \textbf{ground theory term} whose type is \( \typB \) an element of \( \setAlg_\typB \)
by letting \( \mapInt{\appl{\trmA_0}{\trmA_1}}_\typB \) be
\( \mapInt{\trmA_0}_{\typA \arrTyp \typB}(\mapInt{\trmA_1}_\typA) \).
We omit the type and write just \( \mapInt{\cdot} \)
whenever the type can be deduced from the context.
\( \mapInt{\trmA} \) is called the \emph{interpretation} of \( \trmA \).
A \emph{substitution} is a type-preserving map from variables to terms.
Every substitution \( \subA \) extends to a type-preserving map
\( \bar{\subA} \) from terms to terms.
We write \( \trmA \subA \) for \( \bar{\subA}(\trmA) \) and define it as follows:
\( \funA \subA = \funA \),
\( \varA \subA = \subA(\varA) \) and
\( (\appl{\trmA_0}{\trmA_1}) \subA = \appl{(\trmA_0 \subA)}{(\trmA_1 \subA)} \).
Now we postulate a theory sort \( \bSrt \) and
theory symbols \( \type{\bFal}{\bSrt} \) and \( \type{\bTru}{\bSrt} \).
Let \( \setAlg_\bSrt \) be \( \setBra{\bInF,\bInT} \) and
assume \( \mapInt{\bFal} = \bInF \) and \( \mapInt{\bTru} = \bInT \).
A \emph{rewrite rule} \( \rwtr{\rulL}{\rulR}{\rulC} \) is a triple where
\begin{itemize}
\item \( \rulL \) and \( \rulR \) are terms which have the same type,
\item \( \rulL \) is not a theory term,
\item \( \rulC \) is a \emph{logical constraint},
  i.e., \( \rulC \) is a theory term whose type is \( \bSrt \) and
  the type of each variable in \( \setFva{\rulC} \) is a theory sort, and
\item the type of each variable in \( \setFva{\rulR} \setminus \setFva{\rulL} \) is a theory sort.
\end{itemize}
A substitution \( \subA \) is said to \emph{respect} a rewrite rule \( \rwtr{\rulL}{\rulR}{\rulC} \) if
\( \subA(\varA) \) is a value
for all~\( \varA \in \setFva{\rulC} \cup (\setFva{\rulR} \setminus \setFva{\rulL}) \) and
\( \mapInt{\rulC \subA} = \bInT \).
A set \( \setRul \) of rewrite rules induces a \emph{rewrite relation} \( \arrRwt{\setRul} \) on terms such that
\( \trmA \arrRwt{\setRul} \trmA^\prime \) if and only if one of the following conditions is true:
\begin{itemize}
\item \( \trmA = \rulL \subA \) and \( \trmA^\prime = \rulR \subA \)
  for some~\( \rwtr{\rulL}{\rulR}{\rulC} \in \setRul \) and
  some substitution~\( \subA \) that respects \( \rwtr{\rulL}{\rulR}{\rulC} \).
\item \( \trmA = \appl{\funA}{\valA_1 \cdots \valA_n} \) where
  \( \funA \) is a theory symbol but not a value while
  \( \valA_i \) is a value for all~\( i \),
  the type of \( \trmA \) is a theory sort, and
  \( \trmA^\prime \) is the unique value such that
  \( \mapInt{\appl{\funA}{\valA_1 \cdots \valA_n}} = \mapInt{\trmA^\prime} \).
\item \( \trmA = \appl{\trmA_0}{\trmA_1} \),
  \( \trmA^\prime = \appl{{\trmA_0}^\prime}{\trmA_1} \) and
  \( \trmA_0 \arrRwt{\setRul} {\trmA_0}^\prime \).
\item \( \trmA = \appl{\trmA_0}{\trmA_1} \),
  \( \trmA^\prime = \appl{\trmA_0}{{\trmA_1}^\prime} \) and
  \( \trmA_1 \arrRwt{\setRul} {\trmA_1}^\prime \).
\end{itemize}
Logical constraints are essentially first-order---%
higher-order variables are excluded and
theory symbols take only first-order arguments.
We adopt this restriction because
many conditions in functional programs are still first-order and
solving higher-order constraints is hard.
That is not to say that higher-order constraints are of no interest;
we simply leave them out of the scope of LCSTRSs.
Below is an example LCSTRS:
\begin{align*}
  \init                               &\arrRul \appl{\appl{\fact}{n}}{\exit}                                    &&[\bTru] &
  \appl{\appl{\fact}{n}}{k}           &\arrRul \appl{k}{1}                                                      &&[n \le 0]\\
  \appl{\appl{\appl{\comp}{g}}{f}}{x} &\arrRul \appl{g}{(\appl{f}{x})}                                          &&[\bTru] &
  \appl{\appl{\fact}{n}}{k}           &\arrRul \appl{\appl{\fact}{(n - 1)}}{(\appl{\appl{\comp}{k}}{({*}\ n)})} &&[n > 0]
\end{align*}
Here \( \init \) and \( \exit \) denote the start and the end of the program, respectively.
The core of the program is \( \fact \),
which computes the factorial function in continuation-passing style,
and \( \comp \) is an auxiliary function for function composition.
Integer literals and operators are theory symbols and
are interpreted in the standard way.
Note that we use infix notation to improve readability.
The occurrence of \( n \) in the rewrite rule defining \( \init \)
is an example of a variable that
occurs on the right-hand side but not on the left-hand side of a rewrite rule.
Such variables can be used to model user input.
Let \( \setRul \) denote the set of rewrite rules in the example and
consider the rewrite sequence
\begin{align*}
  \appl{\appl{\fact}{1}}{\exit}                                          &\arrRwt{\setRul}
  \appl{\appl{\fact}{(1 - 1)}}{(\appl{\appl{\comp}{\exit}}{({*}\ 1)})} \\&\arrRwt{\setRul}
  \appl{\appl{\fact}{0}}{(\appl{\appl{\comp}{\exit}}{({*}\ 1)})}       \\&\arrRwt{\setRul}
  \appl{\appl{\appl{\comp}{\exit}}{({*}\ 1)}}{1}                       \\&\arrRwt{\setRul}
  \cdots
\end{align*}
In the second step, no rewrite rule is invoked.
Such rewrite steps are called \emph{calculation steps}.
We can write \( \arrRwt{\emptyset} \) for a calculation step.
Terms \( \trmB \) and \( \trmA \) are said to be \emph{joinable} by \( \arrRwt{\emptyset} \),
written as \( \trmB \arrJoi{\emptyset} \trmA \), if
there exists a term \( \trmC \) such that
\( \trmB \arrRwt{\emptyset}^* \trmC \) and
\( \trmA \arrRwt{\emptyset}^* \trmC \).
%

\section{Termination}
In order to prove that a given (unconstrained) TRS \( \setRul \) is terminating,
we usually look for a stable, monotonic and well-founded relation \( \relCwf \)
which orients every rewrite rule in \( \setRul \),
i.e., \( \rulL \relCwf \rulR \) for all~\( \rulL \arrRul \rulR \in \setRul \).
This standard technique, however, requires a few tweaks to be applied to LCSTRSs.
First, stability should be tightly coupled with rule orientation
because every rewrite rule in an LCSTRS is equipped with a logical constraint,
which decides what substitutions are expected when the rewrite rule is invoked.
Therefore, we say that a type-preserving relation \( \relCwf \) on terms
\emph{orients} a rewrite rule \( \rwtr{\rulL}{\rulR}{\rulC} \) if
\( \rulL \subA \relCwf \rulR \subA \) for each substitution~\( \subA \) that \textbf{respects} the rewrite rule.
Second, the monotonicity requirement can be weakened because
\( \rulL \) is never a theory term in a rewrite rule \( \rwtr{\rulL}{\rulR}{\rulC} \).
We say that a type-preserving relation \( \relCwf \) on terms is \emph{rule-monotonic} if
\( \trmA_0 \relCwf {\trmA_0}^\prime \) implies
\( \appl{\trmA_0}{\trmA_1} \relCwf \appl{{\trmA_0}^\prime}{\trmA_1} \) when
\( \trmA_0 \) is not a theory term, and
\( \trmA_1 \relCwf {\trmA_1}^\prime \) implies
\( \appl{\trmA_0}{\trmA_1} \relCwf \appl{\trmA_0}{{\trmA_1}^\prime} \) when
\( \trmA_1 \) is not a theory term.
We present a tentative definition of HORPO \cite{jou:rub:99} on LCSTRSs.
For each theory sort~\( \typA \),
we postulate theory symbols
\( \type{\relTgt_\typA}{\typA \arrTyp \typA \arrTyp \bSrt} \) and
\( \type{\relTge_\typA}{\typA \arrTyp \typA \arrTyp \bSrt} \) such that
\( \mapInt{\relTgt_\typA} \) is a well-founded ordering on \( \setAlg_\typA \) and
\( \mapInt{\relTge_\typA} \) is the reflexive closure of \( \mapInt{\relTgt_\typA} \).
We omit the sort and write just \( \relTgt \) and \( \relTge \)
whenever the sort can be deduced from the context.
Given the \emph{precedence} \( \relPre \), a well-founded ordering on function symbols such that
\( \funA \relPre \funB \) for all~\( \funA \in \setFun \setminus \setThe{\setFun} \) and \( \funB \in \setThe{\setFun} \),
and the \emph{status} \( \mapSta \), a map from \( \setFun \) to \( \setBra{\lblLex,\lblMul_2,\lblMul_3,\ldots} \),
the \emph{higher-order recursive path ordering} (HORPO) \( (\relQua_\rulC,\relCwf_\rulC) \)
is a family of type-preserving relation pairs on terms
indexed by logical constraints and defined as follows:
\begin{itemize}
\item \( \relRpo{\relQua}{\trmB}{\trmA}{\rulC} \) if and only if one of the following conditions is true:
  \begin{itemize}
  \item \( \trmB \) and \( \trmA \) are theory terms whose type is a theory sort,
    \( \setFva{\rulC} \supseteq \setFva{\trmB} \cup \setFva{\trmA} \) and
    \( \rulC \models \appl{\appl{\relTge}{\trmB}}{\trmA} \).
  \item \( \relRpo{\relCwf}{\trmB}{\trmA}{\rulC} \).
  \item \( \trmB \arrJoi{\emptyset} \trmA \).
  \item \( \trmB \) is not a theory term,
    \( \trmB = \appl{\trmB_0}{\trmB_1} \),
    \( \trmA = \appl{\trmA_0}{\trmA_1} \),
    \( \relRpo{\relQua}{\trmB_0}{\trmA_0}{\rulC} \) and
    \( \relRpo{\relQua}{\trmB_1}{\trmA_1}{\rulC} \).
  \end{itemize}
\item \( \relRpo{\relCwf}{\trmB}{\trmA}{\rulC} \) if and only if one of the following conditions is true:
  \begin{itemize}
  \item \( \trmB \) and \( \trmA \) are theory terms whose type is a theory sort,
    \( \setFva{\rulC} \supseteq \setFva{\trmB} \cup \setFva{\trmA} \) and
    \( \rulC \models \appl{\appl{\relTgt}{\trmB}}{\trmA} \).
  \item \( \trmB \) and \( \trmA \) have the same type and
    \( \relRpo{\relUnt}{\trmB}{\trmA}{\rulC} \).
  \item \( \trmB \) is not a theory term,
    \( \trmB = \appl{\funA}{\trmB_1 \cdots \trmB_n} \) where \( \funA \) is a function symbol,
    \( \trmA = \appl{\funA}{\trmA_1 \cdots \trmA_n} \),
    \( \relRpo{\relQua}{\trmB_i}{\trmA_i}{\rulC} \) for all~\( i \) and
    there exists \( i \) such that \( \relRpo{\relCwf}{\trmB_i}{\trmA_i}{\rulC} \).
  \item \( \trmB \) is not a theory term,
    \( \trmB = \appl{\varA}{\trmB_1 \cdots \trmB_n} \) where \( \varA \) is a variable,
    \( \trmA = \appl{\varA}{\trmA_1 \cdots \trmA_n} \),
    \( \relRpo{\relQua}{\trmB_i}{\trmA_i}{\rulC} \) for all~\( i \) and
    there exists \( i \) such that \( \relRpo{\relCwf}{\trmB_i}{\trmA_i}{\rulC} \).
  \end{itemize}
\item \( \relRpo{\relUnt}{\trmB}{\trmA}{\rulC} \) if and only if
  \( \trmB \) is not a theory term,
  \( \trmB = \appl{\funA}{\trmB_1 \cdots \trmB_m} \) where \( \funA \) is a function symbol,
  and one of the following conditions is true:
  \begin{itemize}
  \item \( \relRpo{\relQua}{\trmB_i}{\trmA}{\rulC} \) for some~\( i \).
  \item \( \trmA = \appl{\trmA_0}{\trmA_1 \cdots \trmA_n} \) and
    \( \relRpo{\relUnt}{\trmB}{\trmA_i}{\rulC} \) for all~\( i \).
  \item \( \trmA = \appl{\funB}{\trmA_1 \cdots \trmA_n} \),
    \( \funA \relPre \funB \) and
    \( \relRpo{\relUnt}{\trmB}{\trmA_i}{\rulC} \) for all~\( i \).
  \item \( \trmA = \appl{\funA}{\trmA_1 \cdots \trmA_n} \),
    \( \mapSta(\funA) = \lblLex \),
    \( \relRpo{\relCwf^\lblLex}{\trmB_1 \cdots \trmB_m}{\trmA_1 \cdots \trmA_n}{\rulC} \) and
    \( \relRpo{\relUnt}{\trmB}{\trmA_i}{\rulC} \) for all~\( i \).
  \item \( \trmA = \appl{\funA}{\trmA_1 \cdots \trmA_n} \),
    \( \mapSta(\funA) = \lblMul_k \),
    \( k \le n \),
    \( \relRpo{\relCwf^\lblMul}{\trmB_1 \cdots \trmB_{\min(m,k)}}{\trmA_1 \cdots \trmA_k}{\rulC} \) and
    \( \relRpo{\relUnt}{\trmB}{\trmA_i}{\rulC} \) for all~\( i \).
  \item \( \trmA \) is a value or \( \trmA \in \setFva{\rulC} \).
  \end{itemize}
\end{itemize}
In the above,
\( \relRpo{\relCwf^\lblLex}{\trmB_1 \cdots \trmB_m}{\trmA_1 \cdots \trmA_n}{\rulC} \) if and only if
\( \exists i \le \min(m,n)\ (\relRpo{\relCwf}{\trmB_i}{\trmA_i}{\rulC} \wedge \forall j < i\ \relRpo{\relQua}{\trmB_j}{\trmA_j}{\rulC}) \),
\( \relCwf^\lblMul_\rulC \) is the generalized multiset extension of \( (\relQua_\rulC,\relCwf_\rulC) \) (see \cite{kop:12}),
and \( \rulC \models \rulC^\prime \) denotes,
on the assumption that \( \rulC \) and \( \rulC^\prime \) are logical constraints such that
\( \setFva{\rulC} \supseteq \setFva{\rulC^\prime} \),
that for each substitution~\( \subA \) which maps variables in \( \setFva{\rulC} \) to values,
\( \mapInt{\rulC \subA} = \bInT \) implies \( \mapInt{\rulC^\prime \subA} = \bInT \).
In an unconstrained setting, we make a relation \( \relCwf \) orient a rewrite rule \( \rulL \arrRul \rulR \) by asserting \( \rulL \relCwf \rulR \);
now with the above definition of HORPO on LCSTRSs, if \( \relRpo{\relCwf}{\rulL}{\rulR}{\rulC} \), it is \( \relCwf_\bTru \) that should orient the rewrite rule \( \rwtr{\rulL}{\rulR}{\rulC} \).
Once a combination of \( \relTgt \), \( \relPre \) and \( \mapSta \) that
guarantees \( \relRpo{\relCwf}{\rulL}{\rulR}{\rulC} \)
for all~\( \rwtr{\rulL}{\rulR}{\rulC} \in \setRul \)
is present,
we can conclude that the LCSTRS \( \setRul \) is terminating.
The soundness of this method relies on the following properties of \( \relCwf_\rulC \),
which we must prove:
\begin{itemize}
\item \( \relCwf_\bTru \) orients \( \rwtr{\rulL}{\rulR}{\rulC} \) if
  \( \relRpo{\relCwf}{\rulL}{\rulR}{\rulC} \).
\item \( \relCwf_\bTru \) is rule-monotonic.
\item \( \relCwf_\bTru \) is well-founded.
\item \( \relCom{\arrRwt{\emptyset}}{\relCwf_\bTru} \subseteq {\relCwf_\bTru} \).
\end{itemize}
Note that \( \arrRwt{\emptyset} \) is well-founded because
the size strictly decreases through every calculation step.
Consider the example LCSTRS given in the previous section.
Any combination of \( \relTgt \), \( \relPre \) and \( \mapSta \) that
satisfies the following properties would witness the well-foundedness of \( \arrRwt{\setRul} \):
\( \mapInt{\relTgt} = \lambda x y.\ x > 0 \wedge x > y \),
\( \init \relPre \fact \relPre \comp \),
\( \init \relPre \exit \) and
\( \mapSta(\fact) = \lblLex \).
%

\section{Future Work}
LCSTRSs are still a work in progress.
While the formalism itself is in a somewhat stable state,
the above method for termination analysis is in active development.
First and foremost, we need to prove that HORPO on LCSTRSs has the expected properties.
When the theory is complete, we would like to make a tool to automate the finding of HORPO on LCSTRSs.
It would also be interesting to explore other methods for termination analysis on the new formalism,
including StarHorpo \cite{kop:12} (a transitive variant of HORPO),
interpretation-based methods and dependency pairs.
Another direction is to go beyond LCSTRSs by augmenting the formalism with
lambda abstractions or higher-order constraints.

\bibliography{wst2023}

\end{document}